\documentclass[twocolumn,showpacs,preprintnumbers,amsmath,amssymb]{revtex4}
\usepackage{graphicx}
\usepackage{dcolumn}
\usepackage{bm}

\usepackage{amssymb}

\begin{document}

\title{Localization properties of a tight-binding electronic model \\
on the Apollonian network}
\author{A. L. Cardoso$^1$, R. F. S. Andrade$^{1,2}$, A. M. C. Souza$^3$}
\affiliation{$^1$Instituto de F\'{i}sica, Universidade Federal da Bahia,
40210-210, Salvador, Brazil.\\
$^{2}$Computational Physics, IfB, ETH-H\"{o}nggerberg, Schafmattstr. 6,
8093, Z\"{u}rich, Switzerland.
\\$^{3}$Departamento de F\'{i}sica, Universidade Federal de Sergipe
49.100-000, S\~{a}o Cristov\~{a}o - Brazil.}

\date{\today}

\begin{abstract}

An investigation on the properties of electronic states of a tight-binding
Hamiltonian on the Apollonian network is presented. This structure, which
is defined based on the Apollonian packing problem, has been explored both
as a complex network, and as a substrate, on the top of which physical
models can defined. The Schrodinger equation of the model, which includes
only nearest neighbor interactions, is written in a matrix formulation. In
the uniform case, the resulting Hamiltonian is proportional to the
adjacency matrix of the Apollonian network. The characterization of the
electronic eigenstates is based on the properties of the spectrum, which
is characterized by a very large degeneracy. The $2\pi /3$ rotation
symmetry of the network and large number of equivalent sites are reflected
in all eigenstates, which are classified according to their parity.
Extended and localized states are identified by evaluating the
participation rate. Results for other two non-uniform models on the
Apollonian network are also presented. In one case, interaction is
considered to be dependent of the node degree, while in the other one,
random on-site energies are considered.
\end{abstract}

+

\maketitle

\newpage\


\section{Introduction}

Modeling physical systems on substrates without translation invariance
provides useful insights for the understanding of disordered systems
\cite{Belitz}. Quite recently, the rapid development of complex network
theory \cite{Newmanbook2006} has motivated the investigation of physical
models on such substrates, for the purpose of identifying new aspects that
may stem from novel geometrical constructs \cite{Boccaletti,CostaAdPhys}.
Indeed, the characterization of their physical properties has shown that
new thermodynamic, magnetic and transport features do emerge. In several
situations \cite{Indekeu}, they can become quite distinct from those
obtained by other disordered models.

The Apollonian network \cite{JAndrade}(AN) appears in this scenario as an
interesting structure, in the sense that it shares several properties of
complex networks but, on the other hand, it is defined by strict
geometrical rules. This particular feature makes it possible to explore
scale invariance properties together with numerical methods in the
investigation of distinct models. Its specific geometric features induce
the emergence of non typical behavior in thermodynamic and magnetic
properties of spin systems \cite{RAndrade,Souza}, avalanche distribution
in sand pile models \cite{Vieira}, and so on.

The recent creation of a synthetic nanometer-scale Sierpinski hexagonal
gasket, a self-similar fractal macromolecule \cite{Kome} has highly
motivated the area of scale invariant networks. This perspective raises
the question about electronic systems on these networks.

The purpose of this work is to investigate tight-binding electronic models
on the AN, focusing on the \emph{localized} vs. \emph{extended} nature of
their wave functions. This characterization follows a previous analysis of
the properties of the eigenvalue spectrum of the adjacency matrix (AM) of
the AN \cite{Andrade_spectrum}. The AM spectrum is a very important
network signature and, as such, its properties have been detailed explored
within graph theory. In particular, great attention has been devoted to
understanding iso-spectral graphs that are not isomorphic
\cite{Harary,Dorogovtsev}.

As we will show, the Hamiltonian of the uniform tight-binding model can be
written in terms of the AM network, so that the energy eigenvalues are
proportional to those of the AM. The previous identification of the
structure of the spectrum detailed the presence of several distinct
classes and highly degenerated subsets. The characterization of the
localization properties of the wave function, based on the evaluation of
the participation rate of each state, follows the classification scheme of
different sets in the AM spectrum.

The paper is organized as follows: In Section II, we discuss the
main features of the AN. We also introduce the tight-binding model
and its matrix. We also indicate how the participation rate has been
used to quantify the localization properties of the eigenstates.
Section III brings a brief review of the most important properties
of the uniform eigenvalue spectrum. Results for the three different models
are presented in Section IV. Finally, Section V closes the paper
with a summary of the our main important results.

\section{Tight-binding models on AN}

AN's \cite{JAndrade,Doye} follow from the old problem of optimally filling
a compact domain in the Euclidean plane by circles \cite{Herrmann90}. The
solution to this problem requires the successive addition of maximal
tangent circle into the empty regions bounded by three previously placed
circles. At a given generation $n$, the network $AN_n$ is constructed by
establishing a connection between the centers of all pairs of tangent
circles. The solution to the packing problem requires the determination of
the center and radius of each placed circle. For the AN, however, these
details are not required, and we just need to assign the pairs of nodes
that are connected. In this work, we consider the most simple situation,
which arises by considering, at the zeroth generation $n=0$, three tangent
circles with the same radius, the centers of which occupy the corners of a
equilateral triangle (see Figure 1). The number of sites $N(n)$ on the
network increases according to $N(n)=(3^{n}+5)/2$, while the number of
edges $B(n)$ depends on $n$ as $B(n)=(3^{n+1}+3)/2$. In the limit
$n\rightarrow\infty$, $B(n)/N(n)\rightarrow 3$, so that each node has, in
the average, 6 neighbors.

\begin{figure}
\begin{center}
\includegraphics*[width=4.2cm,height=3.cm,angle=0]{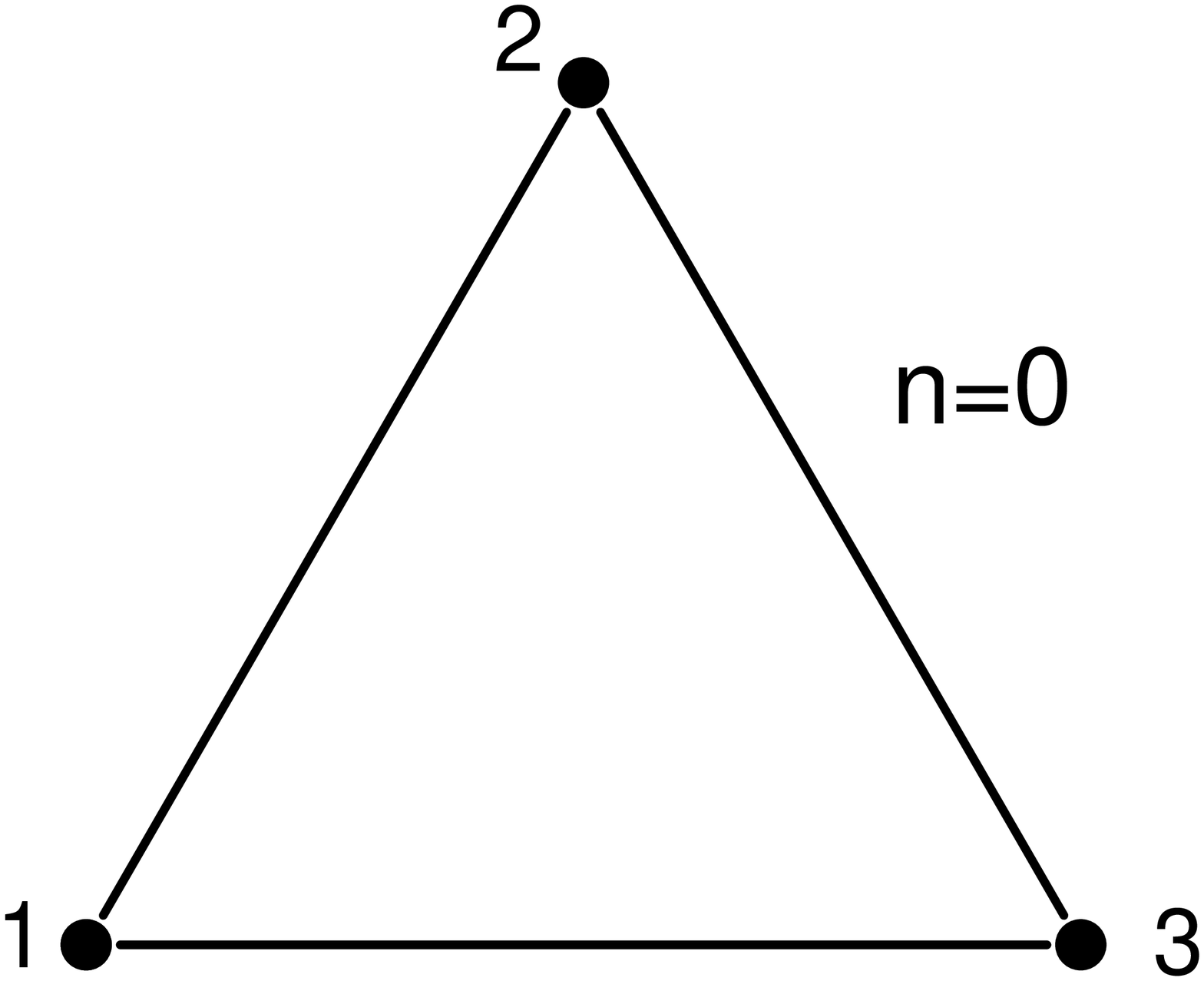}
\includegraphics*[width=4.2cm,height=3.cm,angle=0]{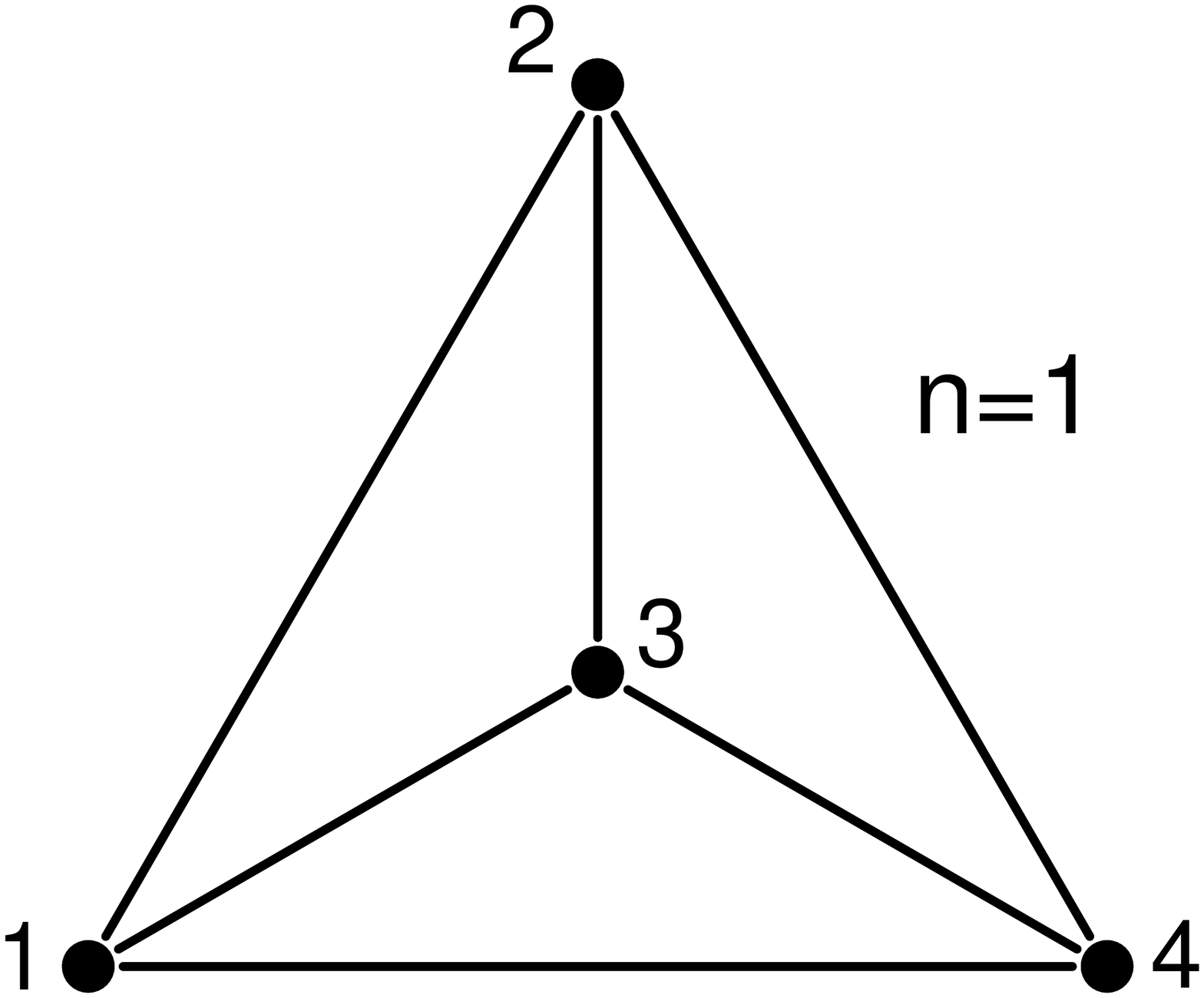}
\includegraphics*[width=4.2cm,height=3.cm,angle=0]{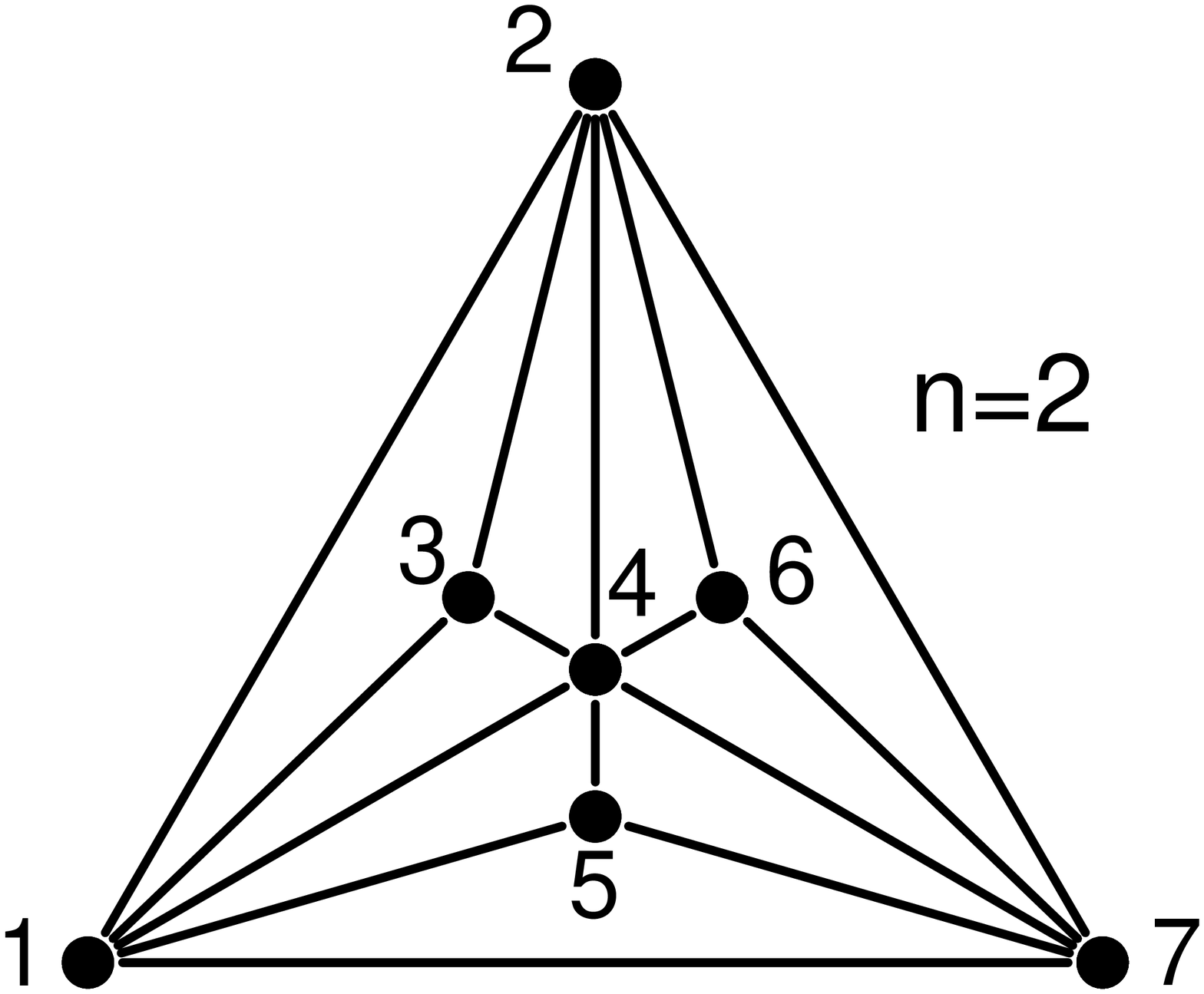}
\end{center}
\caption{Generations $n=0,1,$ and $2$ of AN. The adopted site numbering is
the one introduced in \cite{RAndrade}} \label{fig1}
\end{figure}

The resulting AN shares several properties with other classes of complex
network, including being scale free (the distribution $P(k)$ of node
degree $k$ is a power law), small world (the mean minimal path $\langle
\ell \rangle \sim logN$, and hierarchical (the clustering coefficient of
individual nodes $c(k)$ has a power law dependence on $k$). Nevertheless,
it has own features \cite{CostaAndrade} that precludes being put into the
sets of networks generated by known algorithms, like those proposed by
Watts and Strogatz, \cite{Watts98} and Barabasi and Albert
\cite{Barabasi99}.

The model considers that an electron is primarily bound to a network node $i$
with on-site energy $\epsilon_i$, but it is  also allowed to hop from $i$ to a
neighboring site $j$ one with a probability $V_{i,j}=V_{j,i}$, resulting from
the overlap integral involving the two wave functions and interaction potential.
For a given $n$, the corresponding Hamiltonian is written as

\begin{equation}\label{eq1}
H_n =\sum_{i=1}^N(n) |i\rangle \epsilon_i \langle i | + \sum_{(i,j)}
|i\rangle V_{i,j} \langle j|,
\end{equation}
where $|i\rangle$ represents a Wannier function for an electron at
position $i$, and $(i,j)$ denotes pairs of sites that are first
neighbors.

The Hamiltonian (\ref{eq1}) can be written in a matrix form, where the
diagonal elements $H_{i,i}=\epsilon_i$, while off-diagonal elements
$H_{i,j}$ are set to $V_{i,j}$ or $0$, depending on whether sites $i$ and
$j$ are neighbors or not. The AN sites are numbered according to a
previously introduced scheme, which is reproduced in Figure 1 for $n=0,1,$
and $2$. When we go from generation $n$ to $n+1$, the whole $n-th$ lattice
is first squeezed into the region limited by two outer vertices and the
site at the geometrical center of the triangle. After this first step, two
identical copies of the squeezed lattice, rotated by $2\pi/3$ and
$4\pi/3$, are added to the regions that have been emptied. In this
numbering, two of the outer vertices always take the numbers $1$ and $2$,
while the third one assumes the number $N(n)$ for generation $n$. As a
consequence, the central node receives always the number $N(n-1)$. The
matrix representation of $H_2$ is:

\begin{equation}\label{eq2}
H_2=\left(
  \begin{array}{ccccccc}
    \epsilon_1 & V_{1,2} & V_{1,3} & V_{1,4} & V_{1,5} & 0 & V_{1,7} \\
    V_{2,1} & \epsilon_2 & V_{2,3} & V_{2,4} & 0 & V_{2,6} & V_{2,7} \\
    V_{3,1} & V_{3,2} & \epsilon_3 & V_{3,4} & 0 & 0 & 0 \\
    V_{4,1} & V_{4,2} & V_{4,3}& \epsilon_4 & V_{4,5} & V_{4,6} & V_{4,7} \\
    V_{5,1} & 0       & 0      & V_{5,4} & \epsilon_5 & 0 & V_{5,7} \\
    0       & V_{6,2} & 0      & V_{6,4} & 0 & \epsilon_6 & V_{6,7} \\
    V_{7,1} & V_{7,2} & 0      & V_{7,4} & V_{7,5} & V_{7,6} & \epsilon_7 \\
  \end{array}
\right).
\end{equation}

The simple uniform model corresponds to setting
$\epsilon_i=\epsilon_0,\forall i$, and $V_{i,j}=V_0, \forall (i,j)$. If we
let $E'$ be the eigenvalues of $H$, we can define new energies
$E=(E'-\epsilon_0)/V_0$, which can also be directly obtained by rewriting
(\ref{eq2}), setting $\epsilon_0=0$, and $V_0=1$. In terms of these new
variables, the matrix (\ref{eq2}) coincides exactly with the AM of the AN.
Thus, the eigen-energies of the homogeneous tight-binding model can be
directly obtained from the AM spectrum discussed previously. For the sake
of a simple notation, we will always refer, in the forthcoming discussion,
to the redefined values of $E$. Note, however, that even for the uniform
model, the choice of the on-site energy $\epsilon_0$ can affect the
eigenstate localization properties With the exception of the random
on-site energy model, we restrict ourselves to investigate the situation
in which $\epsilon_0=0$.

Non uniform models can be defined by choosing other values either for the
self-energies or the hopping integrals. In this work we consider a family
of models defined by $\epsilon_i=0$ $\forall i$, and $V(i,j)=1/z^\beta$.
Here, $z=(k_i+k_j)/2$ is the average value of the node degrees $k$, and
the value of $\beta$ selects a particular element of the set. The uniform
model corresponds to setting $\beta=0$. Finally, the third model aims to
investigate the existence of Anderson transition \cite{Anderson} in AN is
defined by setting $V_(i,j)=1$ for neighboring sites, while the
$\epsilon_i$ are chosen according to a probability distribution function.

The solution to the Schr\"{o}dinger equation $H|\Psi_E
\rangle=E|\Psi_E\rangle$ can be written in the basis of Wannier states,
i.e., $|\Psi_E\rangle = \sum_{i=1}^N a_{E,i} |i\rangle$. The site
coefficients $a_{E,i}$ are the components of the normalized eigenvector
corresponding to the eigenvalue $E$, which are obtained from the numerical
diagonalization of the Hamiltonian matrices. The knowledge of the vector
components $a_i$ and the possibility of evaluating it as a function of the
the subsequent generations $n$, suggest the use of the participation ratio
$\xi$ to characterize the degree of localization of the electronic states.
For normalized states, this often used measure is defined by

\begin{equation}\label{eq3}
\xi_{E,n} =\frac{1}{\sum_{i=1}^{N(n)} |a_{E,i}|^4}.
\end{equation}
It can assume values in the range $[1,N]$, which correspond, respectively,
to the extreme situations of completely localized ($a_{E,i}=1$ for one
single value of $i$), and extended states ($a_{E,i}=1/\sqrt{N}, \forall
i$). Thus, it provides both an estimate for the localization length
$\lambda$, as indications for extended states, for which $\xi$ should
scale with $N(n)$. Since the model is defined in the limit
$n\rightarrow\infty$, the localized/extended character of the Apollonian
states of the infinite system can be inferred from the behavior of
$\xi_{E,n}$ as function of $n$.

\section{Spectrum properties}

The evaluation of the eigenvalues must be performed with the help of
numeric algorithms. The spectrum of the uniform model has been
characterized in a detailed way. In this particular case, the presence of
many eigenvalues with high degree of degeneracy, and the recurrent
presence of the same eigenvalues in the successive increasing values of
$n$, makes it possible to present a classification scheme for the
eigenvalues $E$. According to it, for any value of $n\geq2$, we can cast
the eigenvalues into three classes, $C_1^n,C_2^n,C_3^n$. They comprise,
respectively, non-degenerated, two-fold and more than two-fold degenerated
eigenvalues.

A total of $D_1^n=2^{n-1}+1$ and $D_2^n=2^{n}$ eigenvalues are found in
the classes $C_1^n$ and $C_2^n$. Eigenvalues in these classes do not
appear again for any other value of $n$. However, the new eigenvalues
remain in the proximity of the corresponding ones in the former
generation. The total of remaining eigenvalues in the class $C_3^n$ is
$D_3^n=\frac{3}{2}\left(3^{n-1}-2^n+1\right)$. As the total number of
eigenvalues increases with $3^n$, the relative number of states in $C_1^n$
and $C_2^n$ become vanishingly small as $n\rightarrow\infty$.

Contrary to what is observed with the two first classes, eigenvalues in
$C_3^n$ are recurrent, i.e., once they are present in the spectrum for a
given value of $n=q$, they will be found again for any $n>q$. Moreover,
their degeneracy increases with $n$ in a very precise way. Therefore, the
number of new emerging eigenvalues together with their degeneracy can be
quantified according to the following scheme: the number of new degenerate
eigenvalues that emerge at generation $q$ is $2^{q-3}$; at generation $n$,
the degeneracy of any eigenvalue that first appeared at generation $q$ is
$d_{n,q}=\frac{3^{n-q+2}-3}{2}$. Note that $d_{q,q}=3$ always. The first
$C_3$ eigenvalues emerging at $q=3$ and $q=4$ are, respectively, $0$ and
$\pm\sqrt{3}$. The large degeneracy indicates that the characteristic
polynomial is highly factorized. Nevertheless, this knowledge does not
help much in finding exact roots of the polynomials when $q\geq5$.

\section{Results}

\begin{figure}
\begin{center}
\includegraphics*[width=4.2cm,height=3.cm,angle=0]{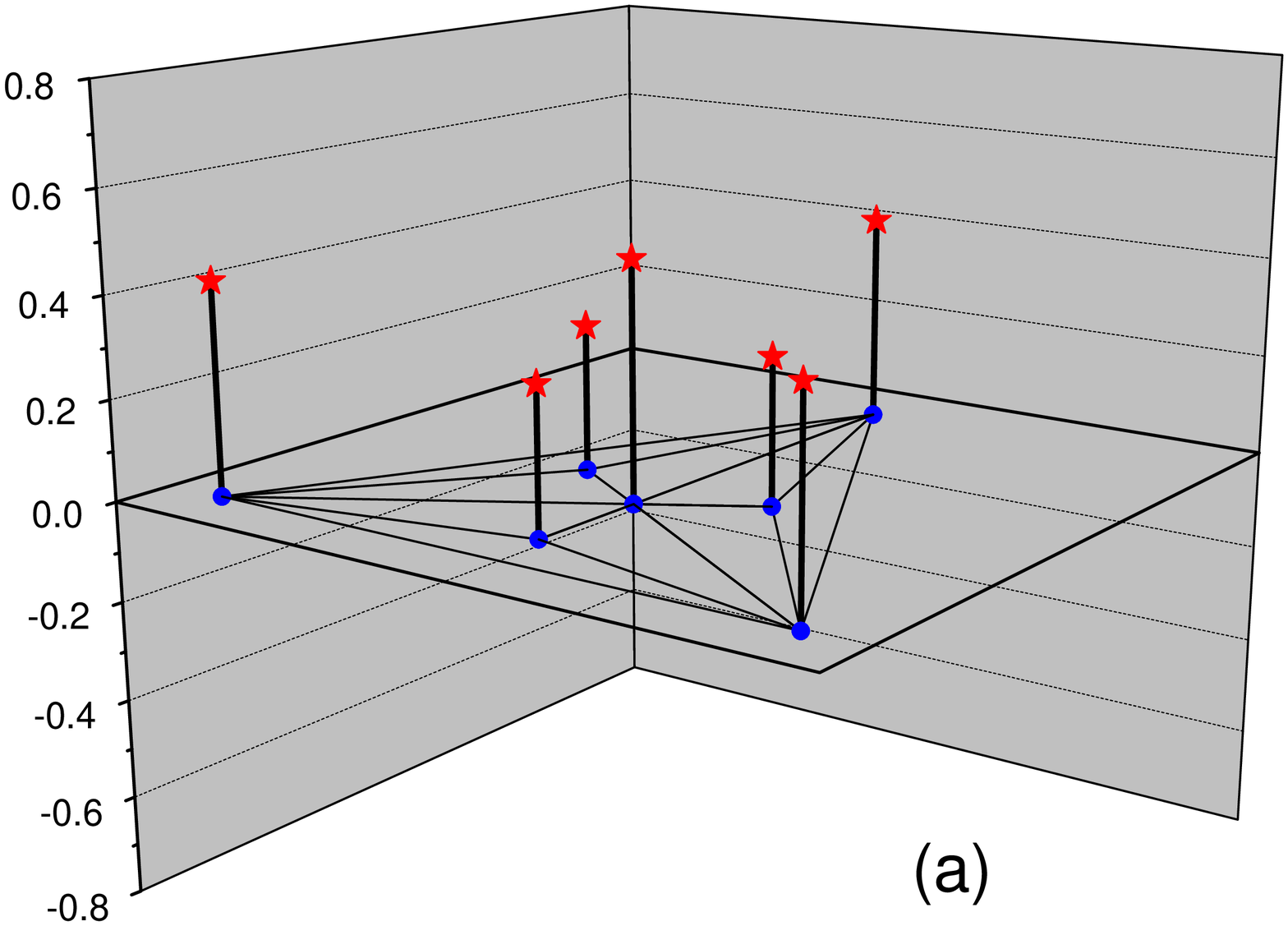}
\includegraphics*[width=4.2cm,height=3.cm,angle=0]{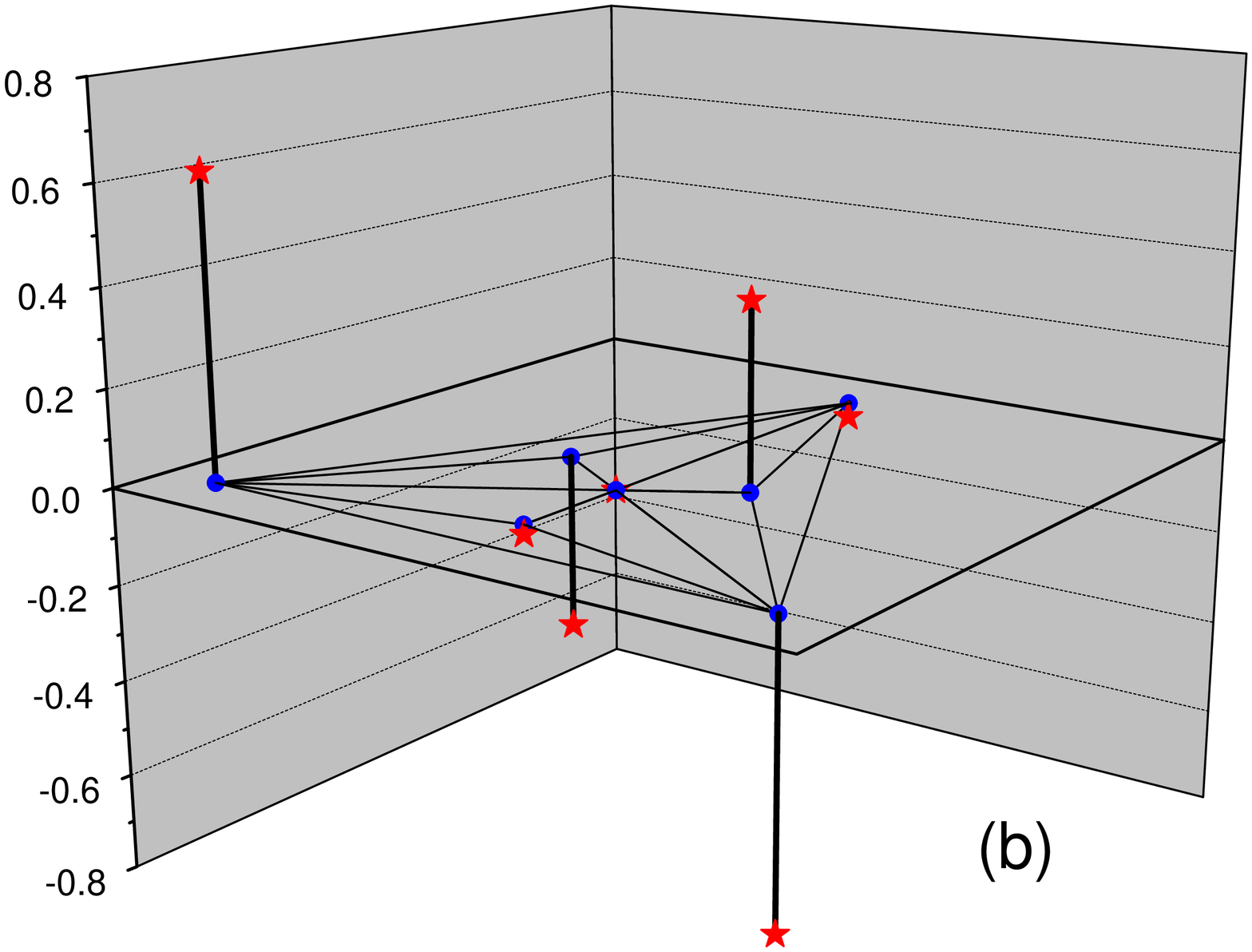}
\includegraphics*[width=4.2cm,height=3.cm,angle=0]{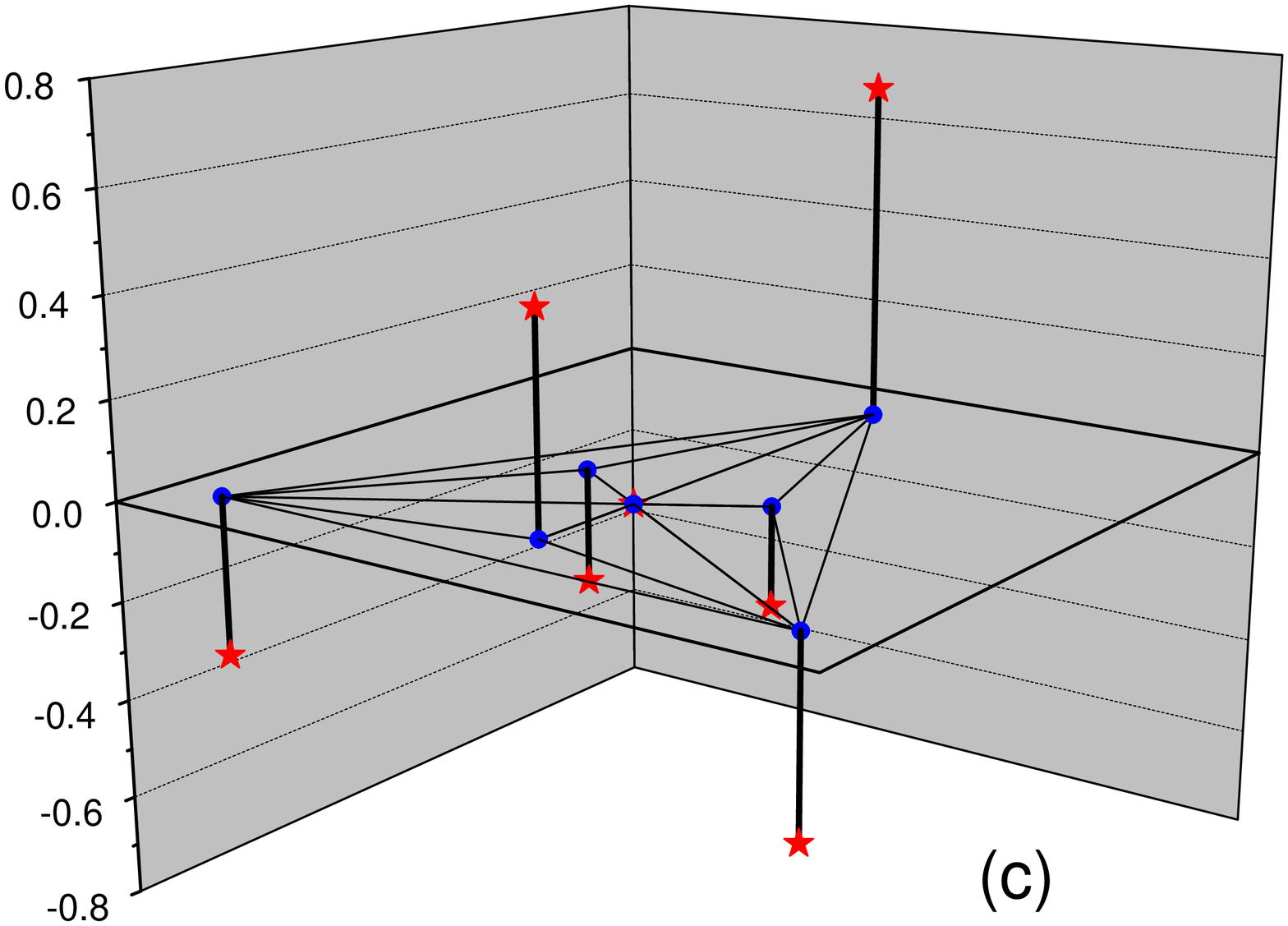}
\end{center}
\caption{Illustration of parity properties for three eigenstates at
generation $n=2$: a) Even state corresponding to the largest value $E$ in
class $C_1^2$; b) and c) Odd states corresponding to a two-fold degenerate
eigenvalue in class $C_2^2$.} \label{fig1}
\end{figure}

\subsection{Symmetry properties}

We have scrutinized all eigenstates until $n=9$, finding  a clear
relationships among several of their properties and the classes to which
the corresponding eigenvalue belongs. The most direct one refers to the
way the states reflect the network invariance by $2\pi/3$ rotations around
the central node. Let us define a parity-like property $P$ based on the
value of  $\sum_{i=1}^{N}a_{E,i}$, so that $P=$ odd (even) when this sum
is zero (non-zero). We have found that, for any  $P=$ even state, all sets
of three equivalent sites, i.e., the three sites that are mapped onto one
another by rotation of the network by $2\pi/3$ and $4\pi/3$ around the
central site, share exactly the same amplitude $a_{E}$. The same amplitude
is found for all sites that are mapped onto one another by inversion
operations over the three bisectrices of the original triangle. All
$C_1^n$ states are even and, conversely, all $C_2^n$ and $C_3^n$ states
are odd.

However, we find two kinds of $P=$ odd states, depending on whether they
belong to the $C_2^n$ or $C_3^n$ classes. Irrespective of the particular
class, the amplitude at the central site for any odd state vanishes
identically, i.e., $(a_{E,i=N(n-1)}=0)$.

In Figure 2 a-c we illustrate some features of three states when $n=2$.
They correspond, respectively, to the largest eigenvalue $E_7$, which
always belongs to $C_n^{(1)}$, and to the second largest two-fold
degenerated eigenvalue $E_5,E_6=-(\sqrt{5}+1)/2$, which is in the class
$C_2^{(2)}$. As exemplary shown for $n=2$, the largest eigenvalue state
has only positive components (absence of nodes). The two states
corresponding to $E_{5,6}$ have odd parity and are clearly orthogonal.

The states in $C_2^n$ are such that, for each set of three equivalent
sites (by rotation of the network by $2\pi/3$ and $4\pi/3$ around the
central site), the sum of the amplitudes $a_i$ is zero. However, no
inversion symmetry, as found for even states, are observed.

The states in $C_3^n$ do not have the same general rotation symmetry
around the central site, but a large number of local $2\pi/3$ rotational
symmetries. Indeed, we found that, if the corresponding eigenvector
appeared for the first time in $C_3^q$, at generation $n$ it contains
$N(n-q+2)$ nodes with zero amplitudes, including the central one. For
instance, when $n=4$, all 12 states corresponding to $E=0$, which appears
for the first time when $n=q=3$, have $N(3)=16$ sites with vanishing
amplitude. The 3 states corresponding to $E=\sqrt{3}$, which appears for
$n=q=4$, have 7 sites with vanishing amplitudes. The sites with vanishing
amplitudes are exactly those sites that are present in the network at
generation $n-q+2$. Further, each set of three sites that appears around
these zero amplitude sites at generation $n-q+3$, have the property that
the sum to the corresponding amplitudes zero. This is illustrated in the
Figure 2d, when $n=q=3$, for a state with eigenvalue $E=0$.

\subsection{Localization properties}

\begin{figure}
\begin{center}
\includegraphics*[width=4.7cm,height=3.3cm,angle=0]{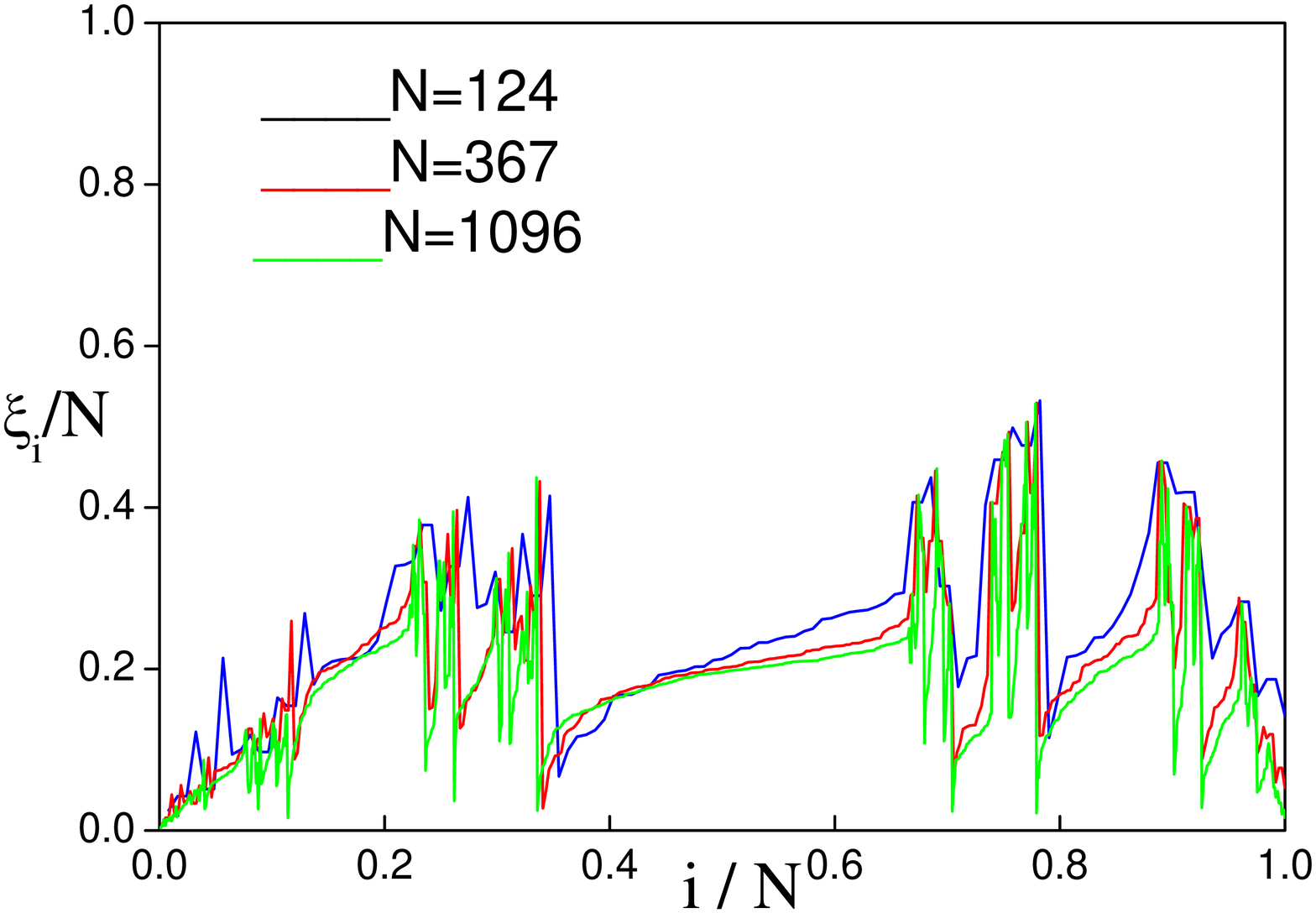}
\end{center}
\caption{Dependence of $\xi_i /N$ with respect to $i/N$ for $n=6,7,8,$ and
$9$. The states are labeled by increasing values of $E$ and, within each
degenerate level, by increasing values of $\xi$.} \label{fig1}
\end{figure}

The localization properties of the quantum states were evaluated with the
help of equation \ref{eq3}. The general features of the results are
displayed in Figure 3 where, in order to better compare the values of
$\xi$ for different values of $n$, we draw the value of $\xi(i)/N(n)$ for
individual states $i$. As the spectrum is highly degenerated, the label
$i$ is chosen to be, in first place, an increasing function of the energy.
Then, within a degenerated level, the states are labeled according to
increasing values of $\xi$. Since the horizontal axis has also been scaled
to $1$ by dividing $i/N$, the graphs show a recurrent form, with the
presence of ever fine details as the value of $n$ increases.

So, despite the shown patterns seem quite irregular, it is possible to
identify large windows where the value of $\xi/N$ increases monotonically.
These structures correspond the highly degenerate levels of $C_3^n$ class,
where the individual states have been classified according to increasing
values of $\xi$. This behavior make it clear that the states corresponding
to degenerate levels have own particular features.

The behavior of the states corresponding to the classes $C_1^n$ and
$C_2^n$ are much difficult to be perceived in Figure 3, as they are
immersed into the overwhelmingly larger number of $C_3^n$ states. However,
the behavior of some of them, at specific positions in the spectrum, can
be identified. For instance, the states corresponding to smallest and
largest value of $E$, both in $C_1^n$ $\forall n$, have very low values of
$\xi$. On the other hand, $C_1^n$ and $C_2^n$ states that precede the
large $C_3^n$ windows, like those just before $E=0$ and $E=\sqrt{3}$, are
characterized by quite large values of $\xi$.

\begin{figure}
\begin{center}
\includegraphics*[width=4.2cm,height=3.cm,angle=0]{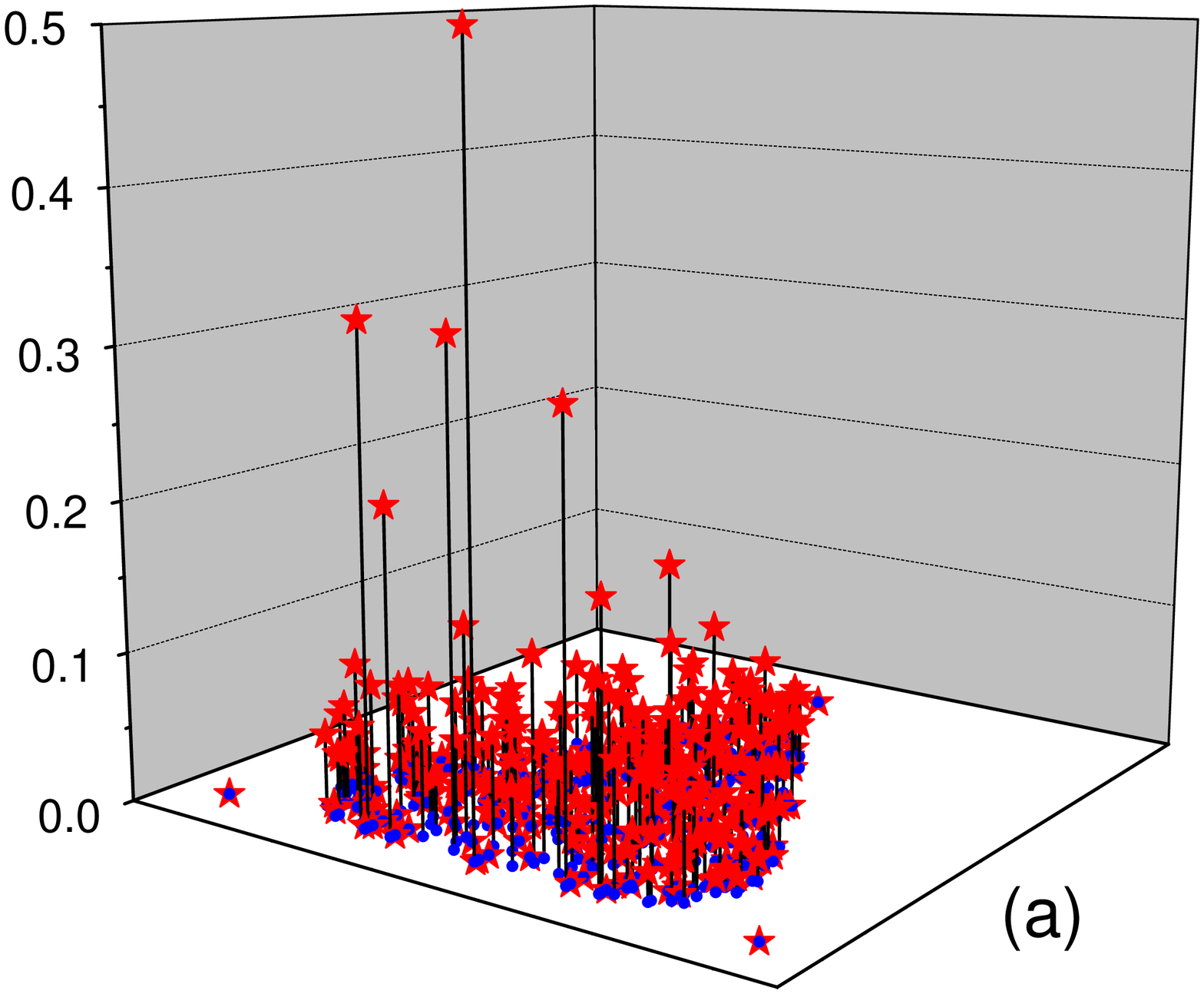}
\includegraphics*[width=4.2cm,height=3.cm,angle=0]{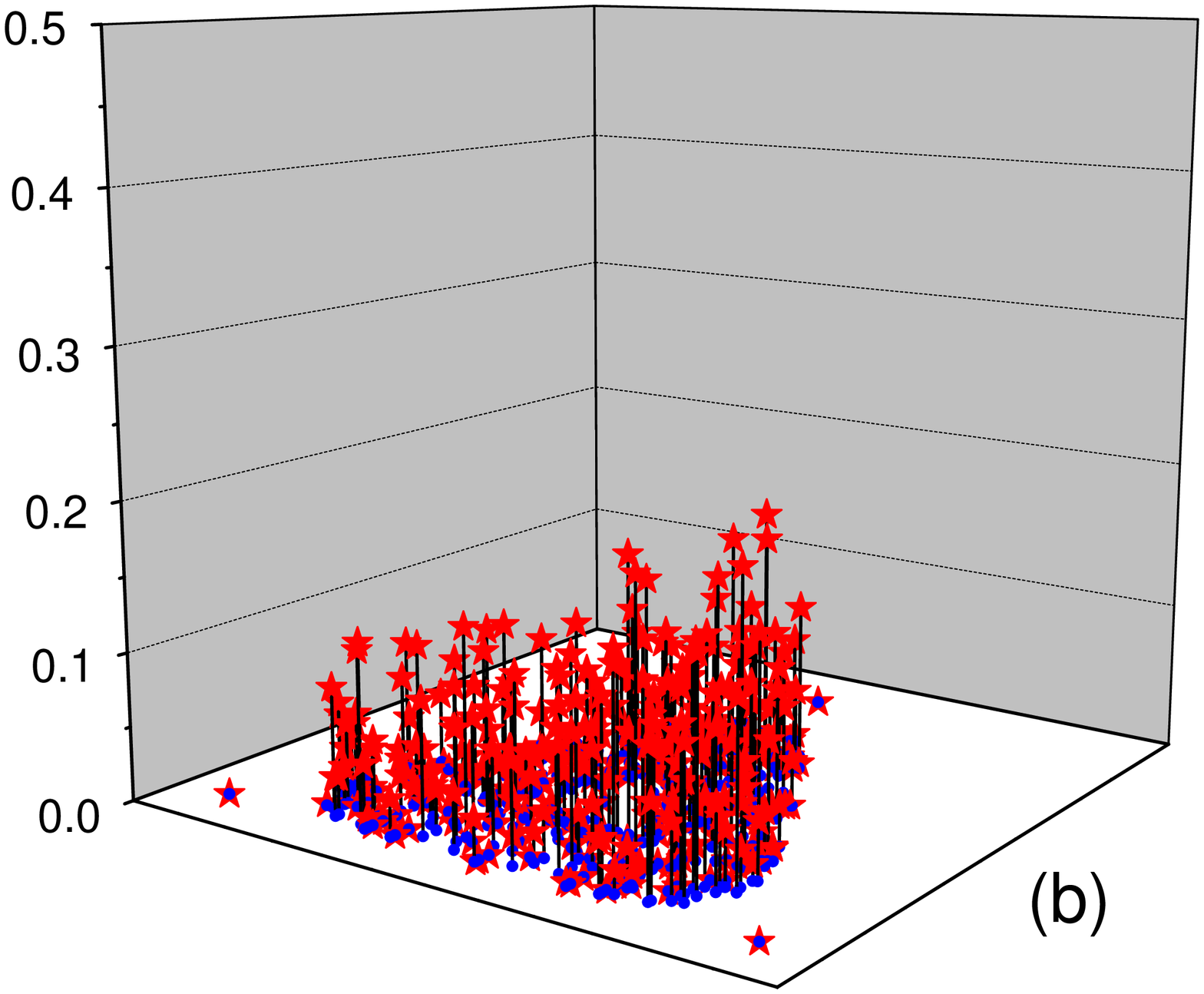}
\includegraphics*[width=4.2cm,height=3.cm,angle=0]{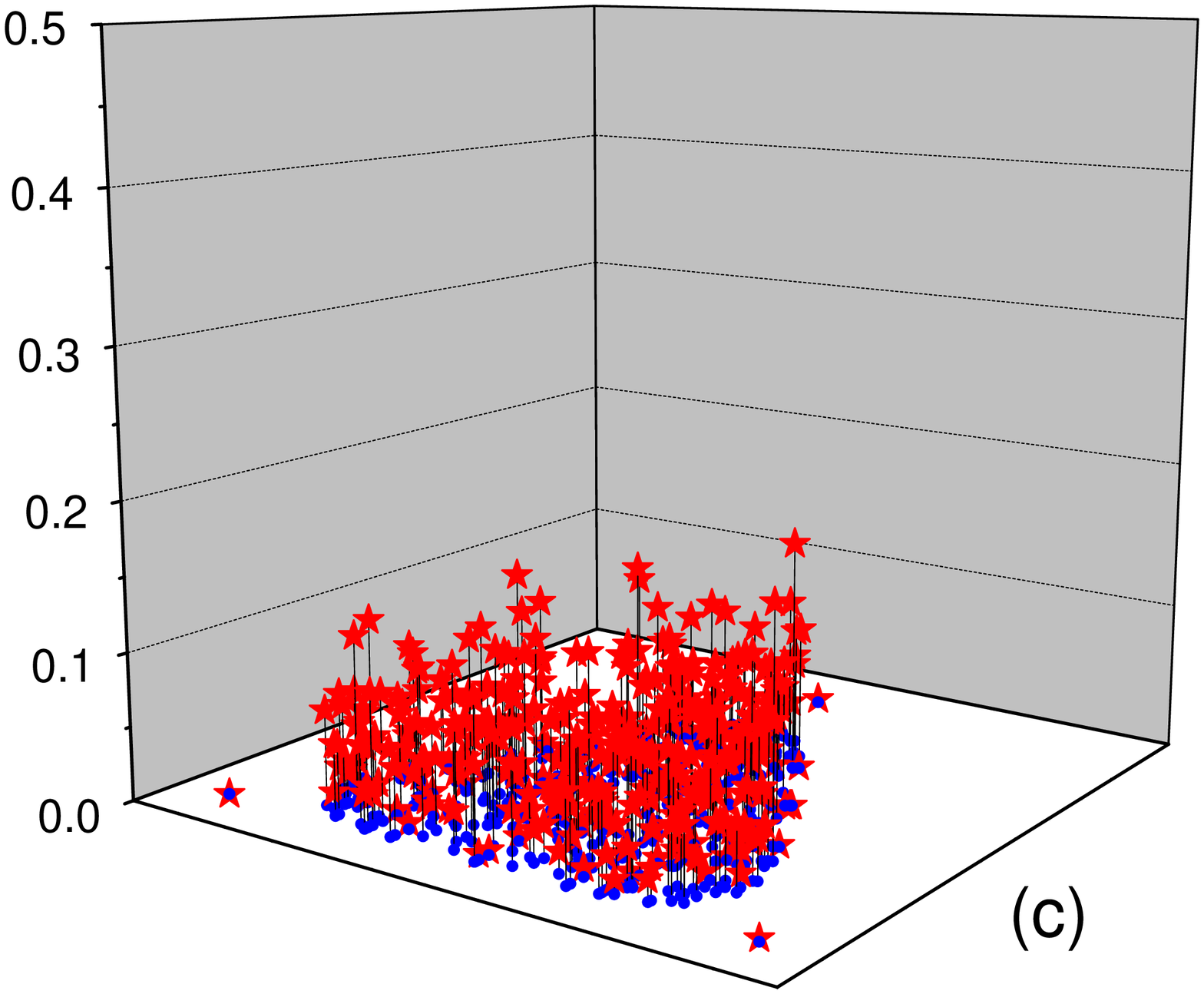}
\end{center}
\caption{Absolute value of wave amplitudes for three states when $n=7$.
(a) and (b) correspond to states with $E=0$ and, respectively, smallest
and largest value of $\xi$. In (c) we show the state with largest value of
$\xi$, in the class  $C_2^7$ and $E=\sqrt{3}$. The increasing value of
$\xi$ stays in close correspondence with the trend to change from
localized to extended character.} \label{fig4}
\end{figure}

In Figure 4a and 4b, we plot the absolute value of local amplitudes for
two $n=7$ states: $i=125$ and $i=244$, both with energy $E=0$ and,
respectively, smallest and largest value of $\xi$. It is possible to
visualize that the state with smallest value of $\xi$ is characterized by
a few large amplitude spikes, distinguish themselves from the much lower
amplitudes in the remaining sites. This picture contrasts with that for
$i=243$, where the amplitudes are homogeneously distributed over all
sites, evidencing properties of an extended state. Figure 4c, for the
$C_2^n$ state $i=286$, with energy $E\simeq=1.0495$, shows a pattern
similar to that in 4b, in agreement with the indication of an extended
state $(\xi/N=0.4181)$.

\begin{figure}
\begin{center}
\includegraphics*[width=4.25cm,height=3.cm,angle=0]{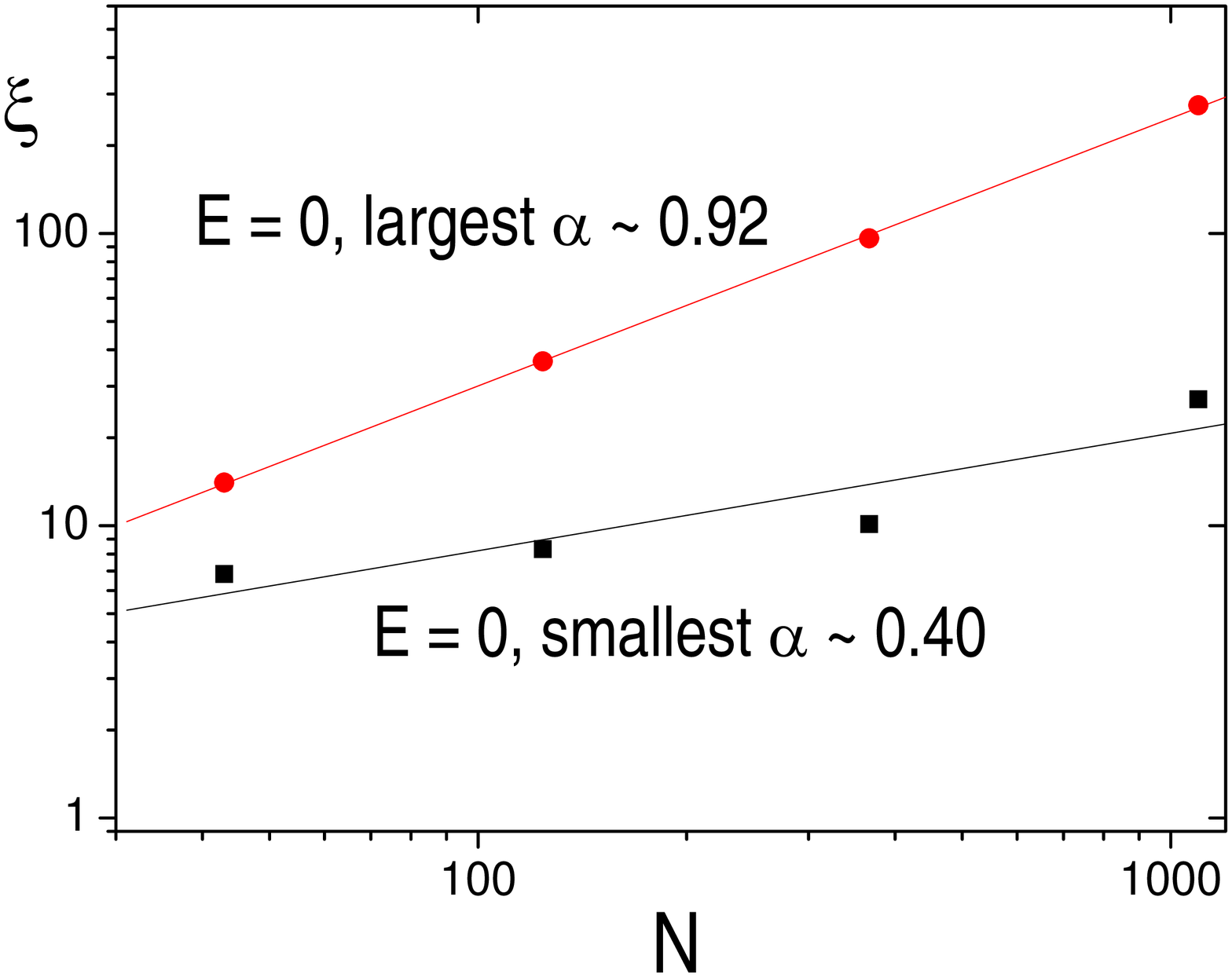}
\includegraphics*[width=4.25cm,height=3.cm,angle=0]{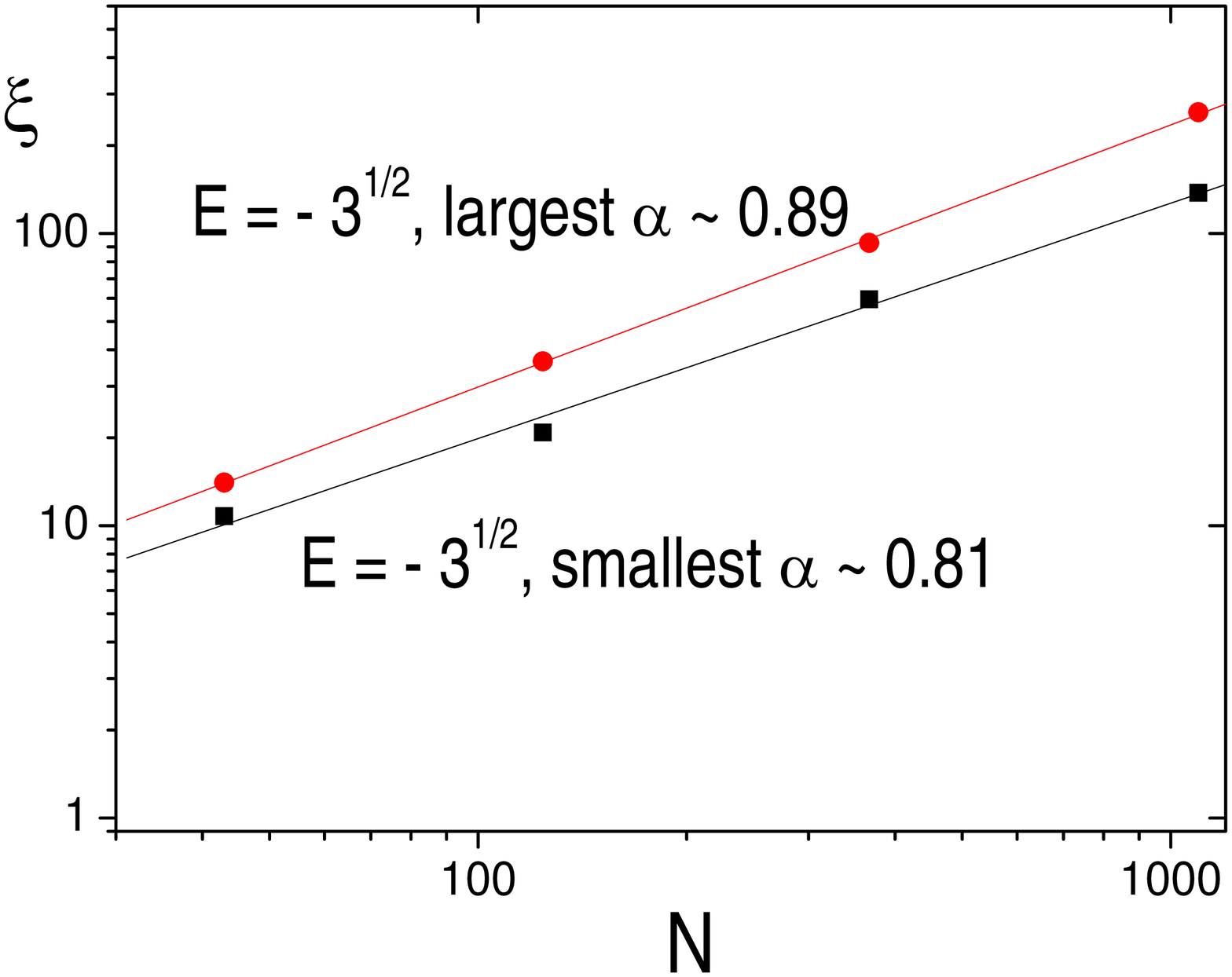}
\includegraphics*[width=4.25cm,height=3.cm,angle=0]{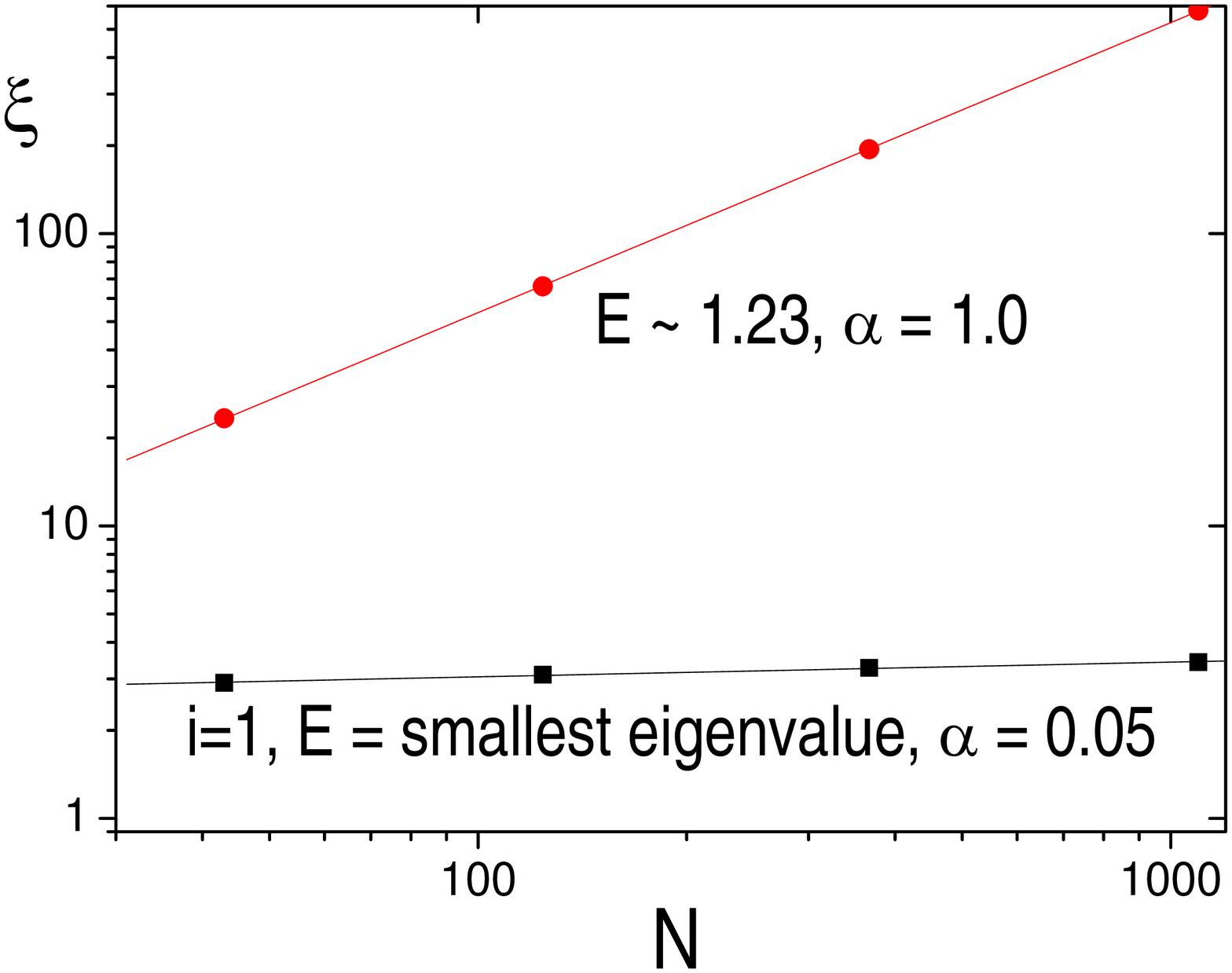}
\includegraphics*[width=4.25cm,height=3.cm,angle=0]{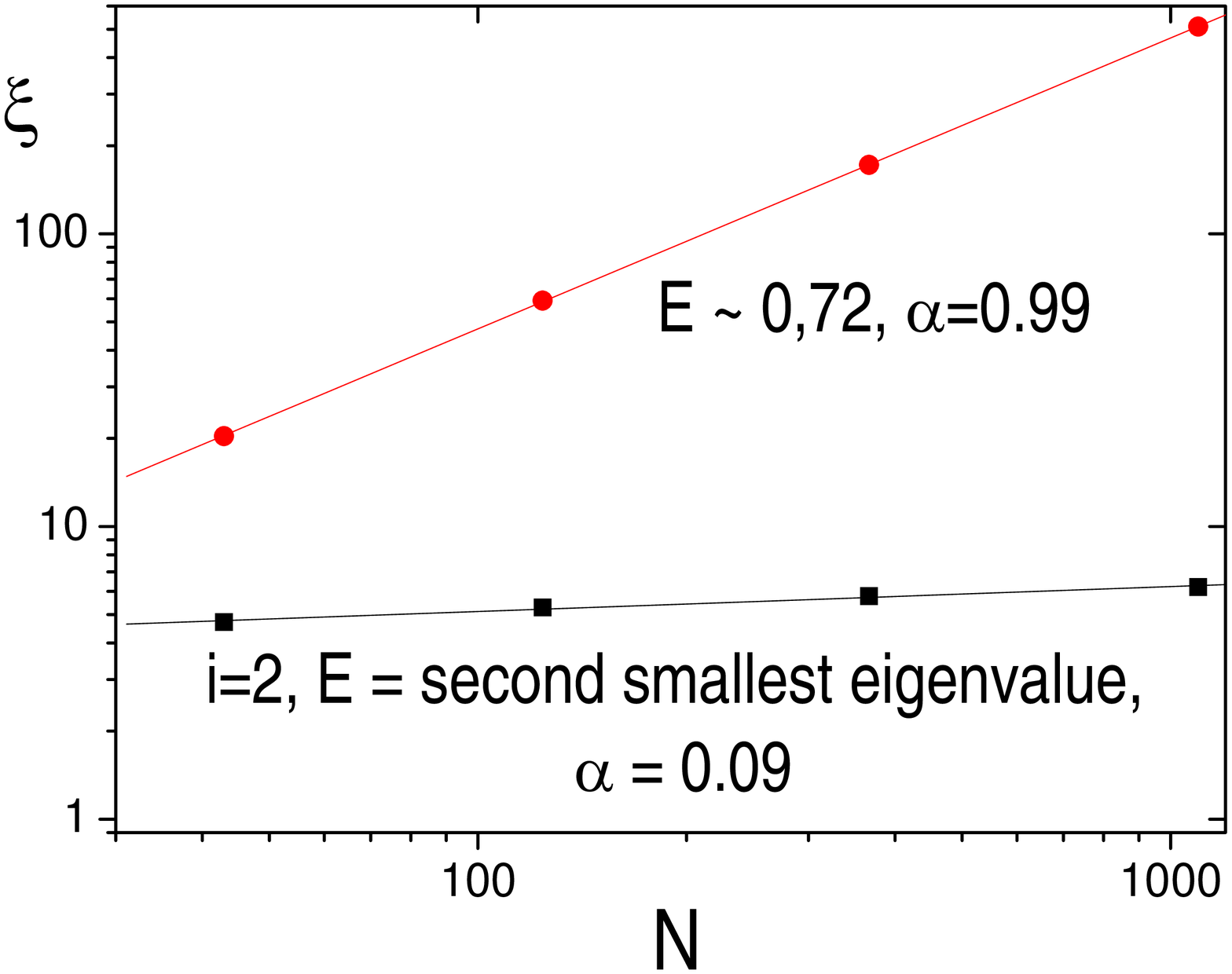}
\end{center}
\caption{Participation ratio $\xi$ for some specific states, as function
of $N$, for generations $n=4,5,6$ and $7$. (a) $E=0$ and (b) $E=-\sqrt{3}$
correspond to states of class $C_3^n$. (c) and (d) illustrate localized
and extended states belonging classes $C_1^n$ and $C_2^n$, respectively.}
\label{fig5}
\end{figure}

To characterize the state properties in the $n\rightarrow\infty$ limit, it
is necessary to follow the behavior of $\xi$ with $n$. As discussed
before, truly extended states obey a linear dependence between $\xi(n)$
and $N(n)$. However, as the number of states increases with the size of
the system, a criterion must be defined to make the correspondence among
the states in different generations $n$. For states in $C_3^n$, such
correspondence is easier to be made, as the eigenvalues in one generation
will be present in all further generations. As this does not happen for
eigenvalues in $C_1^n$ and $C_2^n$, we will compare them by their relative
positions to the eigenvalues in the class $C_3^n$.

In Figure 5, we show the dependence of $\xi$ on the number of sites $N$
for some selected states, from which it is possible to evaluate the
exponent $\alpha$ in $\xi \sim N^\alpha$. In Figure 5a, the two states
with $E=0$ and smallest and largest values of $\xi$ are characterized by
$\alpha=0.40$ and $0.92$, respectively. The indication of localized and
extended character confirm the overall picture shown in Figure 4. The
result in Figure 5b, for the corresponding states of $E=-\sqrt{3}$, reveal
a much smaller variation in the value of $\alpha \in [0.805,0.897]$ for
the states corresponding to extreme values of $\xi$. Similar results are
obtained for other values of $E$ in $C_3^n$, with a clear indication that,
despite the occurrence of some localized states, the results in this class
have a grater tendency of containing states with extended character.

States of the $C_1^n$ class can display both localized and extended
character as indicated by the sequence of values of $\xi$ for the largest
and smallest eigenvalue and for the state with largest value of $\xi$.
While small slopes $\alpha=0.051$ obtained for the sequence of smallest
eigenvalues (Figure 6c) indicate localization, the state with
$E\simeq1.239$, characterized by $\alpha\simeq1$, which maximizes the
value of $\xi$ in all generations, shows the opposite extended pattern.
Similarly, states in class $C_2^n$ can display both localized and extended
character, as shown in Figure 5d. In the first one, the behavior of $\xi$
indicates localized nature for the pair of states corresponding to the
second smallest eigenvalue, while extended properties are obtained for the
pairs of states with $E \simeq 0.72$. In both cases, small vales of
$\alpha$ hints at a strong localization character. On the other hand,
extended properties are found for the series of eigenvalues that precede
the eigenvalue $E=\sqrt{3}$.

\begin{figure}
\begin{center}
\includegraphics*[width=4.2cm,height=3.cm,angle=0]{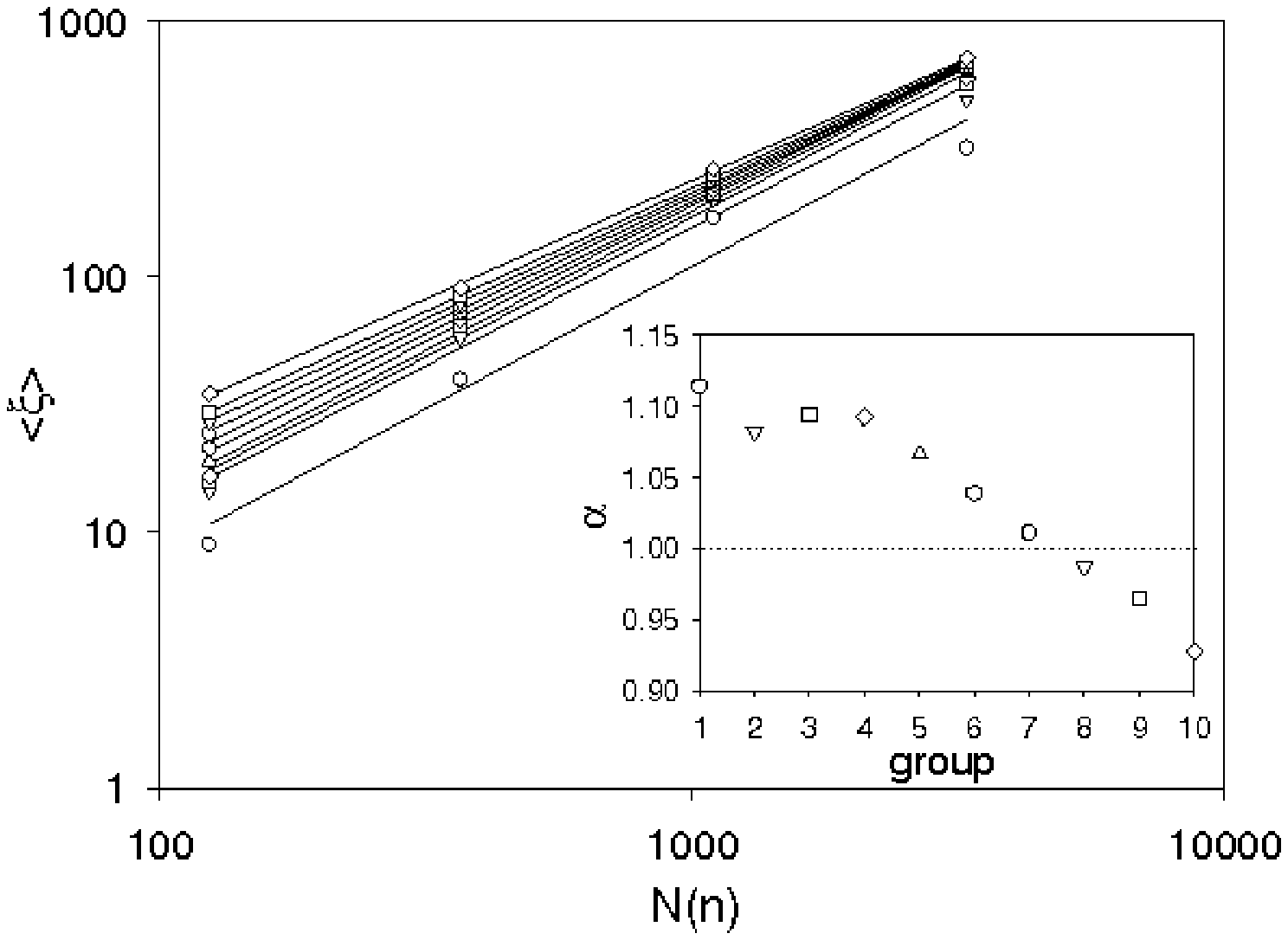}
\includegraphics*[width=4.2cm,height=3.cm,angle=0]{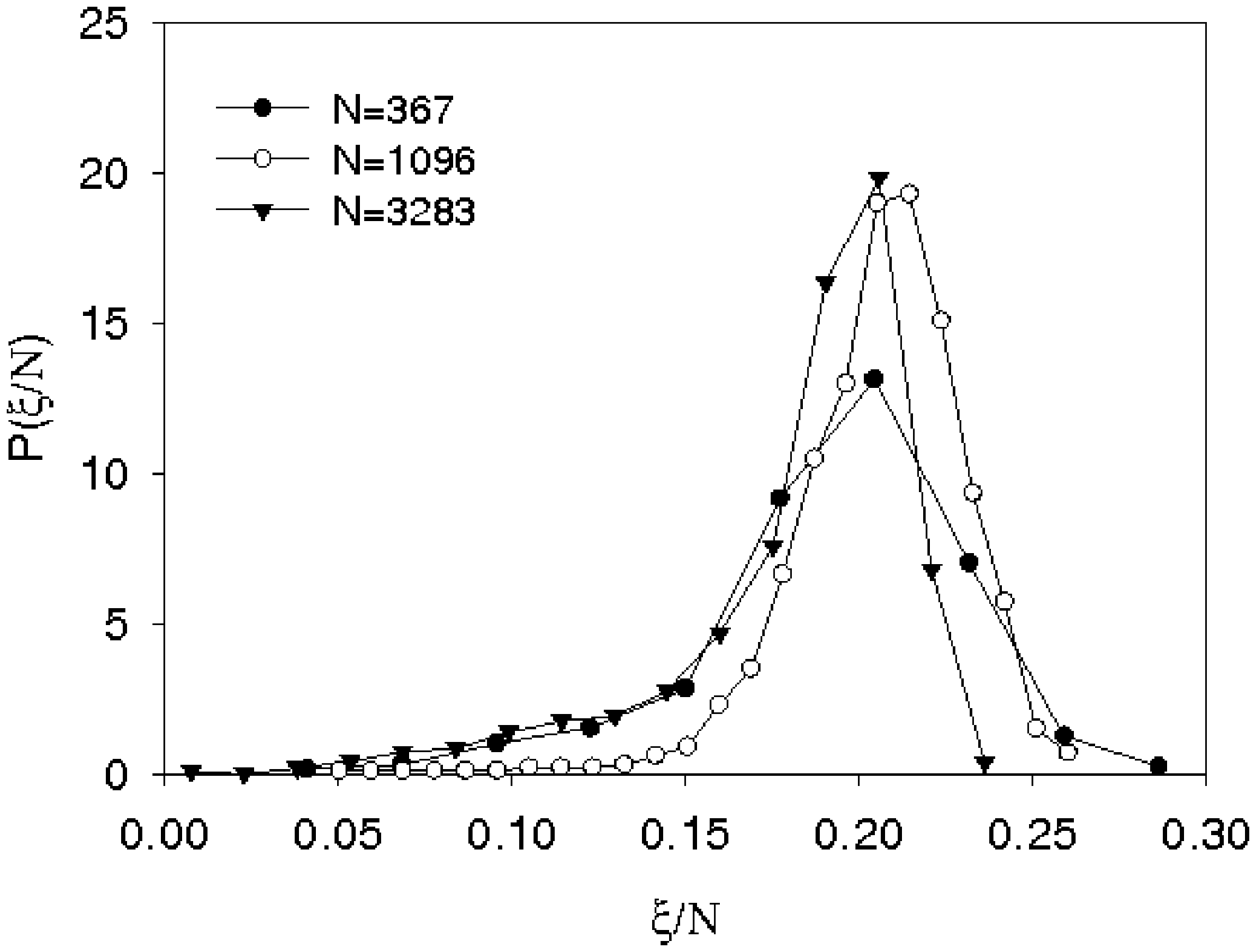}
\caption{(a)Dependence on $N$ of the average participation ratio
$\langle\xi\rangle$, for generations $n=5,6,7$ and $8$. All states with
energy $E=0$ have been casted into 10 groups, according to their values of
$\xi$. The values of the different slopes, drawn in the inset, indicate
that most states are extended. (b) The participation ratio probability
distribution $P(\xi/N)$ of states with $E=0$, for $n=7,8,$ and $9$, is
largely concentrated on large values of $\xi/N$ for all $n$. This is in
accordance with (a).} \label{fig6}
\end{center}
\end{figure}

Since the relative number of states in classes $C_1^n$ and $C_1^n$
decrease with $n$, let us concentrate on the  properties of $C_3^n$
states. A further characterization of their localization character can be
provided by taking the average values of $\xi$ for the states with the
same eigenvalue. For this purpose, for each value of $n$, the states to
each degenerated eigenvalue have be divided into 10 groups, according
increasing values of $\xi$. Then, as in Figure 5, we draw in Figure 6a
$\langle \xi(n) \rangle$ as function of $N(n)$. The values of $\alpha$ for
each subset show, consistently, a tendency to accumulate at a large values
$\geq 1.0$. Another way to confirm the result is to draw the the
participation ratio probability distribution $P(\xi/N)$ as function of
$\xi/N$ (see Figure 6b). For three increasing generations, the position of
the peaks of $P$ indicates that the values of $\xi$ for $E=0$ states
increase linearly with the size of the system, so that their extended
nature is preserved in the limit of infinite system. These features
support our claims of a overwhelming extended character of states in the
uniform model.

\subsection{Localization properties of the non-uniform model}

Let us now discuss the properties of non uniform model mentioned in
Section II, where $V(i,j)=1/z^\beta$. Positive values of $\beta$ decrease
the relative hopping probability from electrons in the largely connected
hubs. So, in principle, we should expect a decrease in the electron
mobility and a tendency for localized states. The opposite happens for
negative values of $\beta$, where the hopping of electrons from the hubs
is enhanced, strengthening the mobility. Since the $V(i,j)$ non-uniformity
destroys the exact scheme based on three well defined classes, it is now
better to provide an analysis of the average behavior of the states,
rather than following them individually.

Our results are summarized in Figure 7. For two distinct values of
$\beta=\pm1$, we qualitatively see, in Figure 7a and 7b, that the
dependence of participation ratio $\xi/N$ with respect to $i/N$ has been
distorted with respect to that of $\beta=0$ (Figure 3). Nevertheless, it
is still possible to identify the position of the $\beta=0$ highly
degenerated $C_3^n$ eigenvalues. In order to quantify the change in the
localization properties as function of $\beta$, we draw in Figure 7c the
behavior of the average value of the state slopes $\langle \alpha \rangle$
with respect to $\beta$. The abrupt change in the average value of
$\langle \alpha \rangle$ at $\beta=3$ hints to an extended-localized
transition of the wave function at this value. This behavior is in
accordance with the discussion in the beginning of this subsection.

\begin{figure}
\begin{center}
\includegraphics*[width=4.2cm,height=3.cm,angle=0]{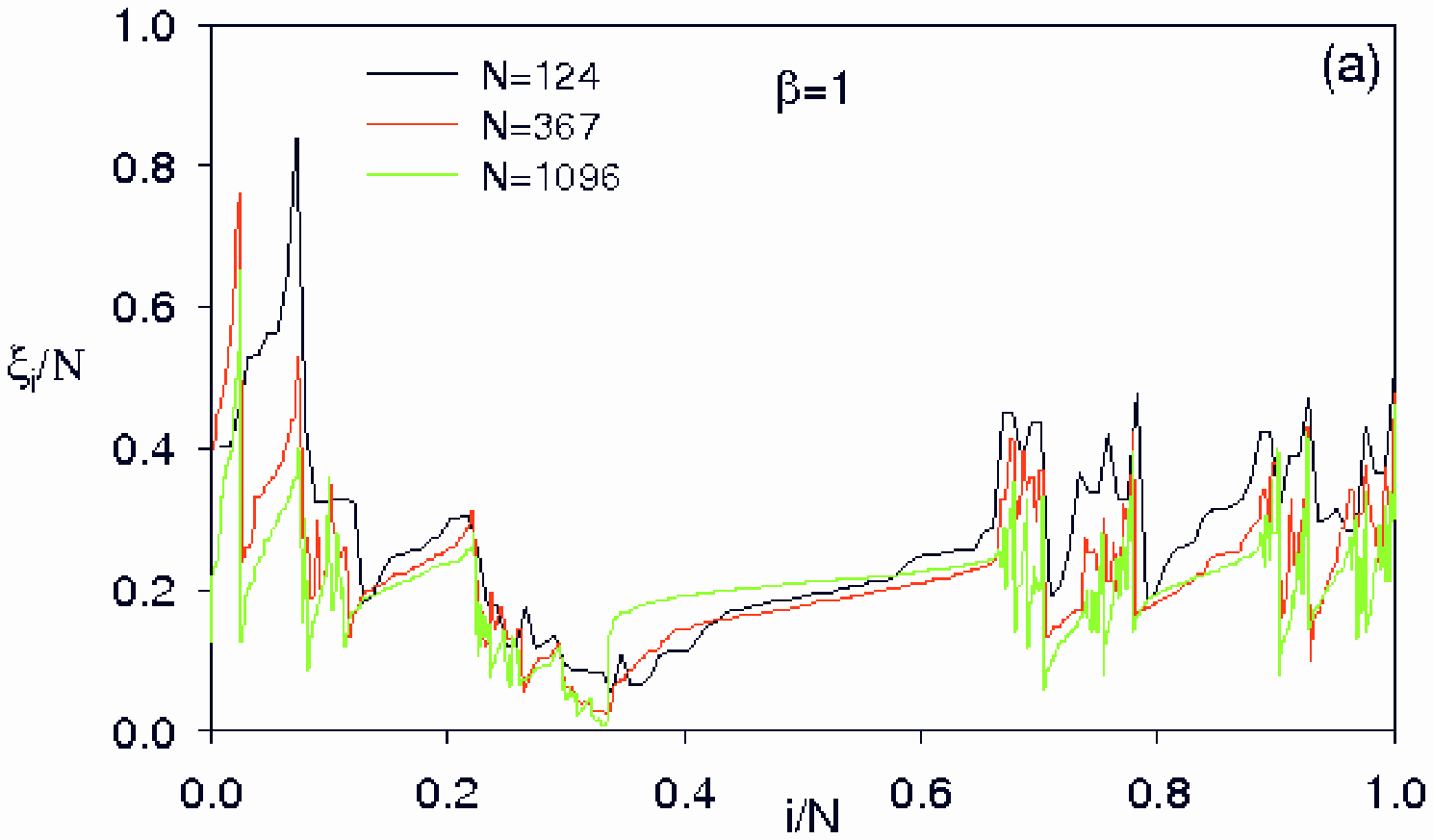}
\includegraphics*[width=4.2cm,height=3.cm,angle=0]{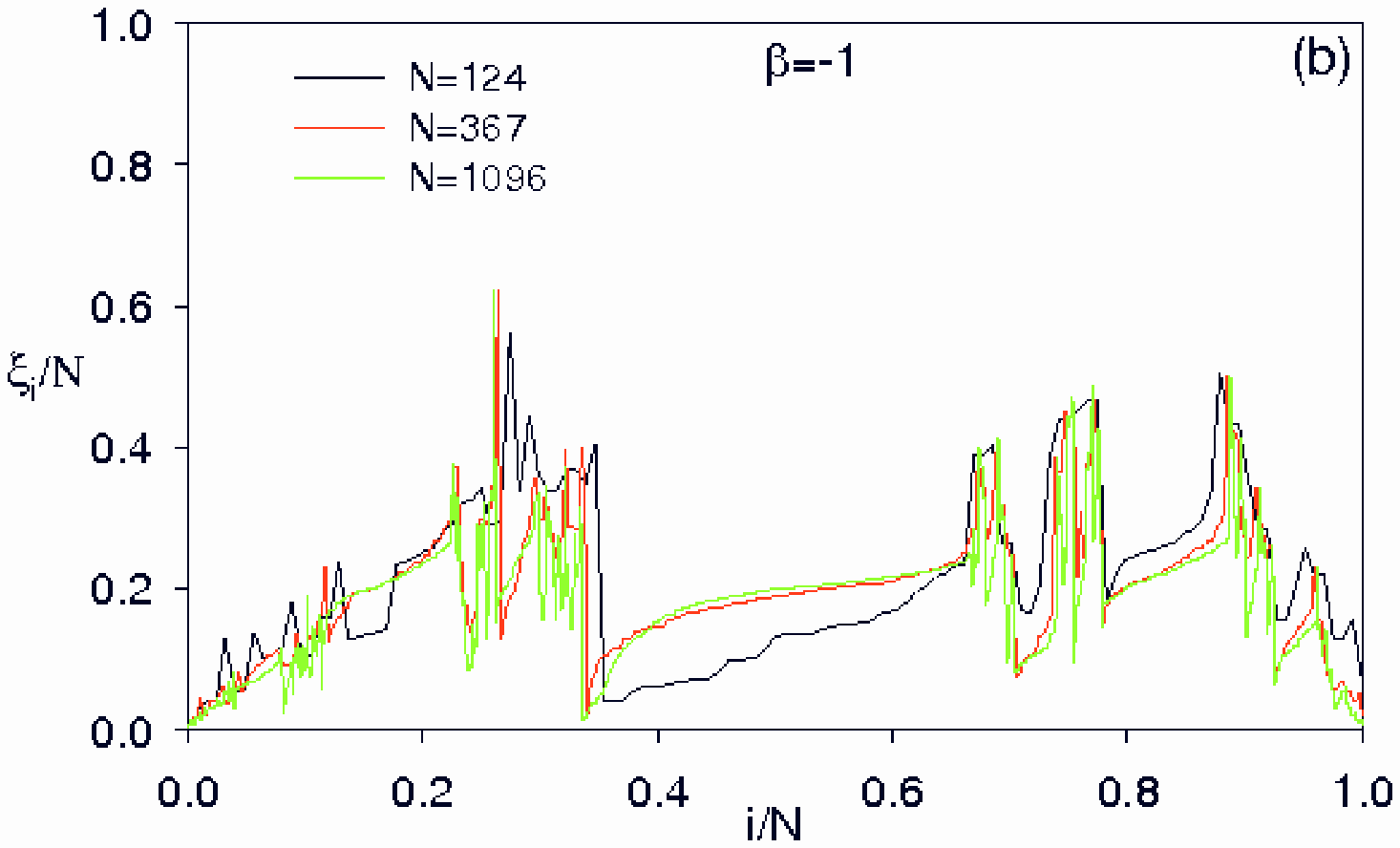}
\includegraphics*[width=4.2cm,height=3.cm,angle=0]{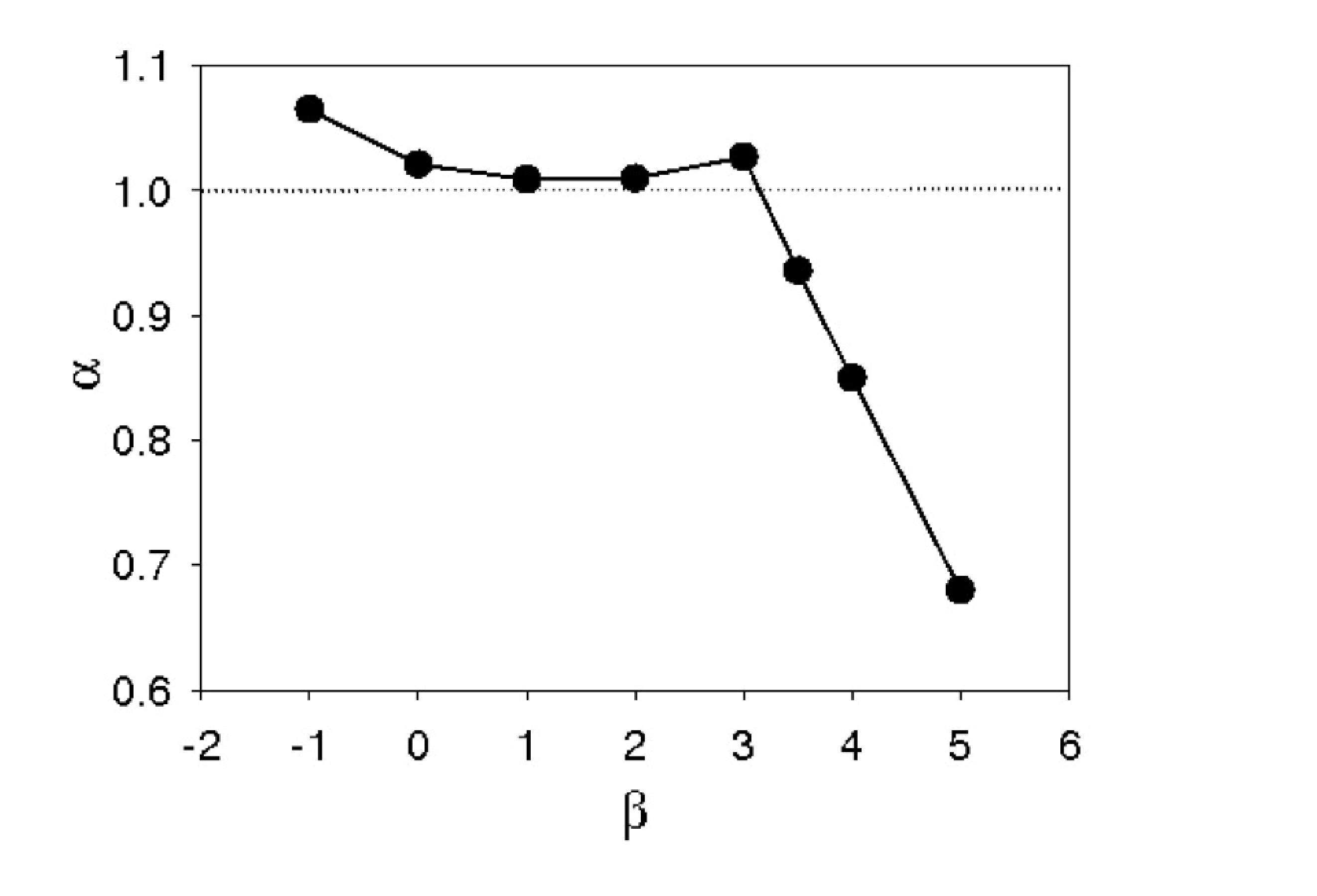}
\end{center}
\caption{Dependence of $\xi_i /N$ with respect to $i/N$ for $n=6,7,$ and
$8$ for $\beta=1$ (a) and $\beta=-1$ (b). The states are labeled by
increasing values of $E$. In (c), we show the dependence of $\langle
\alpha \rangle$ with respect to $\beta$, with a clear indication of
 a sudden change in the nature of the states for $\beta=3$.}\label{fig8}
\end{figure}

\subsection{Anderson transition}
In this subsection we briefly point out the most important features of the
eigenstates that result from introducing disorder into the tight binding
model. In this case, the on-site energy $\epsilon_{i}$ is considered as a
random independent variable, describing the local disorder that disturbs
the motion of electrons. $\epsilon_{i}$ is considered an variable
characterized by a probability function
$P(\epsilon_{i})=\Phi(\Delta/2-|\epsilon_{i}|)/\Delta$, with $\Phi$ being
the step function. The parameter $\Delta$ is a measure for the disorder
strength, and the main purpose is to identify the possible existence of a
Anderson transition for some finite value of $\Delta$.

As observed in the discussion of the non-uniform model, the introduction
of the disorder breaks the high degree of degeneracy of the $C_3^n$
states, as well as the three classes scheme observed in the energy
spectrum of the uniform model. Thus, we resort again to a discussion of
the average properties of the states as a dependency of the energy
interval.

The Anderson transition has been studied taking the average values of
$\xi$ for the all states around energy $|E|\leq \delta E$, with $\delta
E=0.001$. The results are summarized in Table 1, where we present obtained
values of $\langle \alpha \rangle$ as function of $\Delta$. The results
have been obtained by performing $M$ different realizations where,
respectively, $M=1000, 300$ and $50$ for $n=6, 7$ and $8$.

\begin{table}
\begin{center}
\begin{tabular}{|c|c|}
  \hline
  $\Delta$  & $\alpha$ \\ \hline
  0.01 & 0.61 \\
  0.1 & 0.42 \\
  0.2 & 0.38 \\
  0.5 & 0.36 \\
  \hline
\end{tabular}
\caption{Dependence of the localization parameter $\alpha$ on disorder
strength $\Delta$. For all investigated values of $\Delta$, a strong
reduction in the value of $\alpha$ in comparison to the uniform model is
noticed, indicating complete state localization.}
\end{center}
\end{table}

The results in Table 1 can be compared with those obtained for similar
models built on complex network structures. Indeed, Anderson transition
investigations on scale-free networks have shown that the localization of
the electronic states is influenced by the connectivity of the network.
Thus, the fractal dimension quantifies the critical disorder
\cite{sch,sade}. Our investigation confirms that the possibility of the
absence of a Anderson transition at highly connected networks, as observed
in Ref. \cite{sade}.

\section{Conclusions}

In this work we presented a very broad investigation of the properties of
the eigenstates of a tight binding model on the Apollonian network. We
considered the uniform model and two further model versions, where we
could investigate, respectively, the influence of non uniformity induced
by the network geometry and by random on-site energies in the eigenstate
properties.

In the regular case, we could establish a precise relationship between the
three different classes in which the eigenvalue spectrum can be divided
and parity properties of the corresponding eigenvectors. We have shown
that localization property is not related to the eigenvalue classes, by
the identification of eigenstates, with localized and extended properties,
within each of the classes. However, as the number of states in the class
$C_3^n$ increases much faster with $n$ than those of the other two
classes, we proceeded with a quantification of the average participation
ratio of the $C_3^n$ states. The results indicate a scenario in which
extended states dominate the global behavior of the system.

By including an explicit dependence of the node degree on their ability to
decrease or increase the electron mobility, we have shown that the general
state properties can change from the quoted extended character into a
localized one. Our results were based on the ansatz that the interaction
strength decays with the node degree according to a power law. It was
possible to obtain a positive critical value for the change in the
property of the states.

Finally, the model has shown to be much more sensitive to state
localization when random on-site energies are assigned to each site. Our
results suggest that for any non-zero amount of randomness, all states,
even those in the neighborhood of the band center, assume localized
nature.

Physical models constructed on AN's are known to produce quite unusual
properties. These are due mostly on the the existence of large degree
nodes as well as with the existence of large number of loops. Because of
this, it is not possible to affirm that other non-uniform disordered
models on AN will not present other non-expected features.

{\bf Acknowledgement}: This work was partially supported by CNPq. The
authors acknowledge fruitful discussions with Profs. H.J. Herrmann, J.S.
Andrade Jr., and E.L. Albuquerque.

\end{document}